\documentclass[runningheads]{llncs}
\usepackage{graphicx}
\usepackage{subfig}
\usepackage{notoccite}
\usepackage{wrapfig}
\usepackage{amssymb}
\usepackage{cite}
\usepackage{booktabs}
\usepackage[hidelinks,backref=page]{hyperref}
\usepackage[textsize=footnotesize]{todonotes}
\begin{document}

\title{Deep learning for ECoG brain-computer interface: end-to-end vs. hand-crafted features}
%
\titlerunning{End-to-end deep learning for ECoG brain-computer interface}
%
\author{Maciej Śliwowski\inst{1,2}\orcidID{0000-0001-6744-1714} \and
Matthieu Martin\inst{1}\orcidID{0000-0001-5954-8087} \and
Antoine Souloumiac \inst{2} \and
Pierre Blanchart \inst{2} \and
Tetiana Aksenova\inst{1}\orcidID{0000-0003-4007-2343}}

\institute{
Univ. Grenoble Alpes, CEA, LETI, Clinatec, F-38000 Grenoble, France \and
Université Paris-Saclay, CEA, List, F-91120, Palaiseau, France
}
\authorrunning{M. Śliwowski et al.}

\maketitle              
\begin{abstract}
In brain signal processing, deep learning (DL) models have become commonly used. However, the performance gain from using end-to-end DL models compared to conventional ML approaches is usually significant but moderate, typically at the cost of increased computational load and deteriorated explainability. The core idea behind deep learning approaches is scaling the performance with bigger datasets. However, brain signals are temporal data with a low signal-to-noise ratio, uncertain labels, and nonstationary data in time. Those factors may influence the training process and slow down the models' performance improvement. These factors' influence may differ for end-to-end DL model and one using hand-crafted features. 

As not studied before, this paper compares models that use raw ECoG signal and time-frequency features for BCI motor imagery decoding. We investigate whether the current dataset size is a stronger limitation for any models. Finally, obtained filters were compared to identify differences between hand-crafted features and optimized with backpropagation. To compare the effectiveness of both strategies, we used a multilayer perceptron and a mix of convolutional and LSTM layers that were already proved effective in this task. The analysis was performed on the long-term clinical trial database (almost 600 minutes of recordings) of a tetraplegic patient executing motor imagery tasks for 3D hand translation. 

For a given dataset, the results showed that end-to-end training might not be significantly better than the hand-crafted features-based model. The performance gap is reduced with bigger datasets, but considering the increased computational load, end-to-end training may not be profitable for this application.

\keywords{deep learning  \and ECoG \and brain-computer interfaces \and dataset size \and motor imagery \and end-to-end}
\end{abstract}
\section{Introduction}
In the last decade, deep learning (DL) models achieved extraordinary performance in a variety of complex real-life tasks, e.g., computer vision \cite{NIPS2012_krizhevsky}, natural language processing \cite{bert}, compared to previously developed models. This was possible mainly thanks to the improvements of data processing units and, most importantly, increased dataset sizes \cite{NIPS2012_krizhevsky}. Generally, in brain-computer interfaces (BCI) research, access to large databases of brain signals is limited due to the experimental and medical constraints as well as the immensity of paradigms/hardware combinations. Given limited datasets, can we still train end-to-end (E2E) DL models for the medical BCI application as effectively as in computer vision?

In 2019, Roy et al. \cite{Roy_2019} reported that the number of studies classifying EEG signals with deep learning using hand-crafted features (mainly frequency domain) and raw EEG signals (end-to-end) was similar. This indicates that decoding EEG from raw signals is indeed possible. However, in many articles, researchers decided to use harder to design hand-crafted features. While end-to-end models dominated computer vision, in brain signals processing, it is still common to use features extracted as an input to the DL models. It is unclear whether specific signal characteristics cause this, e.g., nonstationarity in time making the creation of a homogeneous dataset impractical, low signal-to-noise ratio complicating the optimization process and favoring overfitting, labels uncertainty originating from human-in-the-loop experimental setup, or researchers' bias toward solutions better understood and more explainable. 

\begin{figure}
\centering
\includegraphics[width=.8\textwidth]{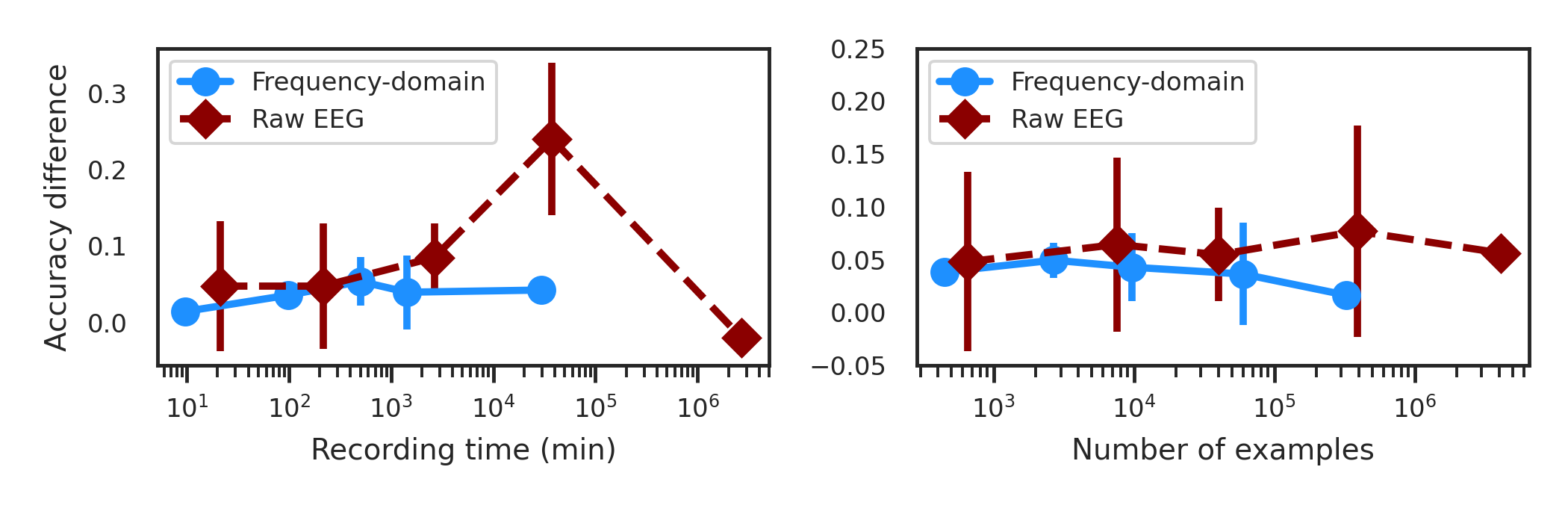}
\caption{Binned average accuracy difference between best proposed DL model and 'traditional' baseline on EEG datasets. Error bars denote one standard deviation of the values in the bin. Bins are equal in size on a logarithmic scale. Points x-axis position denotes the average dataset size in a bin.} \label{fig:intro-acc-vs-dataset-size}
\end{figure}

Most studies do not directly compare DL using end-to-end and hand-crafted features approaches. Usually, DL architectures are compared with each other and with an additional 'traditional' ML pipeline, e.g., filter-bank common spatial pattern (FBCSP) in \cite{schirrmesiter_eeg}, xDAWN and FBCSP in \cite{Lawhern_2018}, SVM and FBCSP in \cite{Tabar_2016}. In figure \ref{fig:intro-acc-vs-dataset-size}, we presented accuracy improvement of the best proposed DL model compared to the 'traditional' baseline for articles analyzed in \cite{Roy_2019} \footnote{limited to the articles that contained all the required information, code adapted from \cite{Roy_2019}} depending on the recording time and the number of examples in the dataset. The gap between performance improvement of DL compared to the 'traditional' baseline increases with the dataset size (except for the last points on the plot, which contain significantly fewer studies). In the right plot, the difference between models using raw EEG and frequency domain features increases which may exhibit a boost of end-to-end models with access to bigger datasets compared to hand-crafted features. As the proposed DL models are usually compared to the baseline, the boost of end-to-end models cannot be clearly stated because the accuracy difference depends strongly on the 'traditional' baseline model performance and the particular task tackled in the study. 

While EEG and ECoG signals share many characteristics---both are multichannel temporal signals with information encoded in frequency and space, with low signal-to-noise ratio and noisy labels---there are also differences, e.g., a higher spatial resolution of ECoG, higher signal-to-noise ratio and higher contribution of informative high gamma band ($>70$Hz). In motor imagery ECoG decoding, end-to-end DL is not commonly used. 'Traditional' ML classifiers are usually preceded by a feature extraction step creating brain signals representation, typically in the form of time-frequency features, containing information about power time course in several frequency bands \cite{liang_decoding_2012,schalk_decoding_2007} or focused only on low-frequency component (LFC)/Local
Motor Potential (LMP) \cite{schalk_decoding_2007} (detailed analysis can be found in \cite{volkova_ecog_review}).

However, a successful application of an end-to-end DL model to motor imagery decoding of finger movements trajectory from ECoG was performed with convolutional layers filtering the raw signal both in temporal and spatial domains followed by LSTM layers \cite{xie_decoding_2018}. Smart weights initialization was helpful in achieving high performance. Nevertheless, an average improvement from training the weights can be estimated as $0.022 \pm 0.0393$ of Pearson r correlation coefficient, which is relatively small, with 66\% of cases noticeable improvement from end-to-end training (at the level of subjects/fingers). As this was not studied before, we investigated the differences in data requirements between an end-to-end model and one using hand-crafted features on a long-term clinical trial BCI dataset of 3D target reach task. Unique long-term recordings (several months of experiments, more than 600 min duration in total, compared to few minutes of ECoG recording available in previous studies, e.g., \cite{xie_decoding_2018}) allowed us to explore the relationship between dataset size and the type of feature used for ECoG signal decoding. In this study, we used architectures previously applied to the ECoG dataset for decoding motor imagery signals with hand-crafted time-frequency features as input \cite{sliwowski_2022}. In addition, we optimized the temporal filtering layer with backpropagation seeking a more efficient set of filters that were initialized to reproduce continuous wavelet transform. We also investigated whether both approaches react differently to training dataset perturbations which may be the case due to distinct model properties and may influence the choice of optimal data processing pipeline for ECoG BCI.

\section{Methods}

\subsection{Dataset}

The dataset used in this study was collected as a part of the clinical trial 'BCI and Tetraplegia' (ClinicalTrials.gov identifier: NCT02550522, details in \cite{benabid_exoskeleton_2019}) approved by the ethical Committee for the Protection of Individuals (Comité de Protection des Personnes—CPP) with the registration number: 15-CHUG-19 and the Agency for the Safety of Medicines and Health Products (Agence nationale de sécurité du médicament et des produits de santé---ANSM) with the registration number: 2015-A00650-49
and the ethical Committee for the Protection of Individuals (Comité de Protection des Personnes---CPP) with the registration number: 15-CHUG-19.

\begin{wrapfigure}[19]{r}{0.5\textwidth}
    \centering
    \includegraphics[width=.5\textwidth]{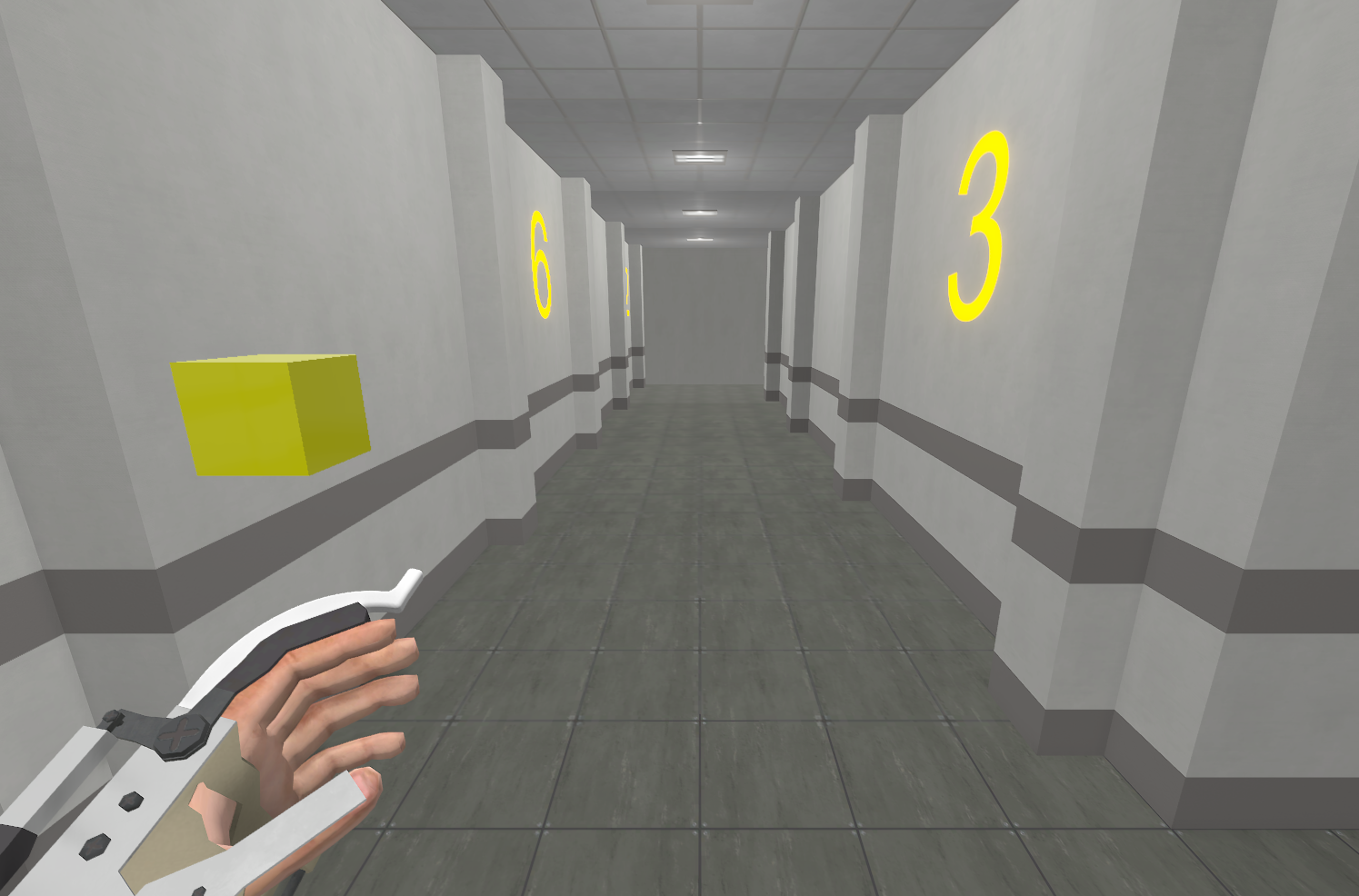}
    \caption{Screenshot from the virtual environment. The patient is asked to reach the yellow square (target) with the left hand (effector) using motor imagery.}
    \label{fig:virtual-environment}
\end{wrapfigure}

In the experiment, a 28-years-old tetraplegic patient after spinal cord injury was asked to move the hands of a virtual avatar displayed on a screen (see figure \ref{fig:virtual-environment}) using motor imagery patterns---by imaging/attempting hand movements that influence brain activity in the motor cortex. These changes were then recorded with two WIMAGINE \cite{mestais_wimagine_2015} implants placed over the primary motor and sensory cortex bilaterally. Both implants consisted of $8 \times 8$ grid of electrodes with recording performed using 32 electrodes selected in a chessboard-like manner due to limited data transfer with a sampling frequency equal to 586 Hz. Signals from implants were transferred to the decoding system that performed online predictions. First, one out of 5 possible states (idle, left and right hand translation, left and right wrist rotation) was selected with a state decoder. Then, for every state (except idle), a multilinear REW-NPLS model \cite{eliseyev_recursive_2017} updated online was used to predict 3D movements or 1D wrist rotation. The dataset consisted of 44 experimental sessions recorded over more than 200 days. It constitutes 300 and 284 minutes for left and right hand translation, respectively.

\subsection{Data representation and problem}
Based on the collected database, we extracted two datasets for left and right hand translation. The raw signal representation was created from 1-second long windows of ECoG signal with 90\% overlap. Every observation $\mathbf{X}_i \in \mathbb{R}^{64 \times 590}$ contained 590 samples for each of the 64 channels corresponding to the number of electrodes recording the signal.

Every signal window $\mathbf{X}_i$ was paired with the corresponding desired trajectory $\mathbf{y}_i \in \mathbb{R}^{3}$ that the patient was asked to follow, i.e., the straight line connecting the tip of the hand to the target. The trajectories were computed in the 3D virtual avatar coordinate system mounted in the pelvis of the effector.

Before feeding the data to the models, datasets were cleaned from data loss artifacts that were not caught during the online recordings. Additionally, observations for which the predicted and desired state did not match due to state decoder errors were also removed to reduce the number of mislabelled observations (e.g., when the patient was asked to control left hand translation but instead left wrist was rotating).

Then, all the models were trained to find the mapping between $\mathbf{X}_i$ ECoG signal and $\mathbf{y}_i$ desired trajectories that the exoskeleton hand should follow in the case of optimal prediction. As a performance metric we used cosine similarity (equation \ref{eq:cs}) measuring cosine of the angle $\alpha_i$ between prediction $\hat{\mathbf{y}}_i$ and the desired trajectory $\mathbf{y}_i$.

\begin{equation}
	\textrm{CS}(\mathbf{y}_i, \hat{\mathbf{y}}_i) = \frac{\mathbf{y}_i \cdot \hat{\mathbf{y}}_i}{\|\mathbf{y}_i\| \cdot \|\hat{\mathbf{y}}_i\|} = \cos{\alpha_i}
	\label{eq:cs}
\end{equation}
Cosine loss defined as $\textrm{CL}(\mathbf{y}_i, \hat{\mathbf{y}}_i) = 1 - \textrm{CS}(\mathbf{y}_i, \hat{\mathbf{y}}_i)$ was used as optimization objective.

\subsection{'Traditional' feature extraction and DL optimization}
Hand-crafted features were extracted using complex continuous wavelet transform (CWT). CWT was performed with Morlet wavelets with central frequencies ranging from 10 to 150 Hz with a step of 10 Hz. Each wavelet support consisted of 118 samples (0.2s) centered on its maximum value. Features were obtained by applying CWT on one-second-long signals, computing the module of the complex signals, and performing an average pooling of 0.1 second. The resulting feature tensor was of shape $64 \times 15 \times 10$, with dimensions corresponding to channels, frequency bands, and time steps.

CWT can be represented as a convolution between a set of filters and a signal in the temporal domain. In the standard case, the filters are fixed and constitute a basis for feature extraction where every filter detects brain activity in a different frequency band. As every spatial channel is convolved separately in time, we obtained a time-frequency-space representation of the ECoG signal (see table \ref{tab:feature_extractor} for feature extractor architecture specification).

Here, we propose to adjust the filters during backpropagation together with all other parameters of the models. In the first scenario, the filters were initialized to Morlet wavelets with 15 central frequencies, resulting in 30 kernels (real and imaginary parts). Note that at the beginning of training, the first layer reproduces 'traditional' feature extraction. The filters were fixed for 5 epochs of so-called pre-training, then they were unfreezed and optimized for the following 50 epochs. The pre-training was used to not distort the wavelets drastically in the first epochs when parameters of the rest of the network are randomly initialized. We also evaluated random weights initialization from uniform distribution as a solution that does not incorporate prior knowledge about the system. 

In the second scenario, an alternative approach was used to maintain the wavelet structure by optimizing only the parameters used to generate the wavelets instead of modifying all filters' parameters. In our case, the function generating the wavelets was defined as: 
\begin{equation}
    \Psi(t, f) = \frac{1}{\sqrt{\pi}} \frac{1}{\sqrt{\frac{f_s}{f}}} e^{-(t f)^2} e^{2i\pi t f}
\end{equation}
where central frequency parameter $f$ defines the center of the frequency band analyzed by the wavelet and $f_s$ is the signal sampling frequency. In the central frequency optimization (CFO) scenario, we optimized only the central frequency $f$ parameters (one per wavelet), so the filters after training are still from the Morlet wavelets family.


\begin{table}
\caption{The architecture used to reproduce hand-crafted feature extraction with CWT. Only one convolutional layer (conv time) was used in computations according to the performed experiment E2E/E2E CFO.} \label{tab:feature_extractor}
\centering
\begin{tabular}{@{}lllll@{}}
\toprule
Layer                  & Kernel Shape           & Output Shape             & Param \# & Mult-Adds      \\ \midrule
Input                  & --                     & {[}200, 1, 590, 8, 8{]}  & --       & --             \\
\hspace{5mm} Conv time              & {[}1, 30, 118, 1, 1{]} & {[}200, 30, 590, 8, 8{]} & 3,570  & 27,006,336,000 \\
\hspace{5mm} Conv time CFO          & {[}1, 30, 118, 1, 1{]} & {[}200, 30, 590, 8, 8{]} & 15       & 27,006,336,000 \\
Square                 & --                     & {[}200, 30, 590, 8, 8{]} & --       & --             \\
Sum real and imaginary & --                     & {[}200, 15, 590, 8, 8{]} & --       & --             \\
Square root            & --                     & {[}200, 15, 590, 8, 8{]} & --       & --             \\
Dropout                & --                     & {[}200, 15, 590, 8, 8{]} & --       & --             \\
AvgPool                & --                     & {[}200, 15, 10, 8, 8{]}  & --       & --             \\
BatchNorm              & {[}15{]}               & {[}200, 15, 10, 8, 8{]}  & 30       & 6,000          \\ \bottomrule
\end{tabular}%
\end{table}

\subsection{DL architectures}
In this study, we used two architectures proposed in \cite{sliwowski_2022}, i.e., CNN+LSMT+MT, which showed the best performance, and MLP, which was the simplest approach. In the baseline approach, the 'traditional' feature extraction was followed with fully connected or convolutional layers. When optimizing the first convolutional layer, we kept the rest of the network the same to isolate the influence of the training feature extraction step. Details of the tested DL architectures are described below and in \cite{sliwowski_2022}. Additionally, we used ShallowFBCSPNet and Deep4Net \cite{schirrmesiter_eeg} as end-to-end DL baseline.

\subsubsection{MLP}
The most basic DL architecture evaluated in the study was multilayer perceptron (MLP), consisting of two fully connected layers. Dropout and batch normalization layers were placed between fully connected layers for stronger regularization (see table \ref{tab:mlp}).

\begin{table}
\caption{MLP architecture from \cite{sliwowski_2022}.} \label{tab:mlp}
\centering
\begin{tabular}{@{}lllll@{}}
\toprule
Layer           & Kernel Shape   & Output Shape    & Param \# & Mult-Adds  \\ \midrule
Flatten         & --             & {[}200, 9600{]} & --       & --         \\
Fully connected & {[}9600, 50{]} & {[}200, 50{]}   & 480,050  & 96,010,000 \\
BatchNorm       & {[}50{]}       & {[}200, 50{]}   & 100      & 20,000     \\
ReLU            & --             & {[}200, 50{]}   & --       & --         \\
Dropout         & --             & {[}200, 50{]}   & --       & --         \\
Fully connected & {[}50, 50{]}   & {[}200, 50{]}   & 2,550    & 510,000    \\
ReLU            & --             & {[}200, 50{]}   & --       & --         \\
Dropout         & --             & {[}200, 50{]}   & --       & --         \\
Fully connected & {[}50, 3{]}    & {[}200, 3{]}    & 153      & 30,600           \\ \bottomrule
\end{tabular}%
\end{table}

\subsubsection{CNN+LSTM+MT}
In the CNN+LSTM+MT architecture, CWT features were further analyzed with $3 \times 3$ convolutional layers in space (electrodes organized on an array $4 \times 8$ reflecting positions of electrodes on implants). After two convolutional layers, two LSTM layers were applied to analyze temporal information from 10 timesteps. Finally, every output of the last LSTM layer was used for training to compute loss based on all predicted and ground truth trajectories corresponding to 1 second (10 timesteps) of signal analyzed (see table \ref{tab:cnn}).

\begin{table}
\caption{CNN+LSTM+MT architecture from \cite{sliwowski_2022}.} \label{tab:cnn}
\centering
\begin{tabular}{@{}lllll@{}}
\toprule
Layer                         & Kernel Shape      & Output Shape        & Param \# & Mult-Adds   \\ \midrule
Input                         &                   & [200, 15, 8, 8, 10] & --      &             \\
\hspace{5mm}Input per implant &                   & [200, 15, 8, 4, 10] & --      &             \\
\hspace{5mm}Conv space        & [15, 32, 3, 3, 1] & [200, 32, 6, 4, 10] & 4,352   & 208,896,000 \\
\hspace{5mm}ReLU              & --                & [200, 32, 6, 4, 10] & --      & --          \\
\hspace{5mm}BatchNorm         & [32]              & [200, 32, 6, 4, 10] & 64      & 12,800      \\
\hspace{5mm}Dropout           & --                & [200, 32, 6, 4, 10] & --      & --          \\
\hspace{5mm}Conv space        & [32, 64, 3, 3, 1] & [200, 64, 4, 2, 10] & 18,496  & 295,936,000 \\
\hspace{5mm}ReLU              & --                & [200, 64, 4, 2, 10] & --      & --          \\
\hspace{5mm}Dropout           & --                & [200, 64, 4, 2, 10] & --      & --          \\
LSTM                          & --                & [200, 10, 50]       & 215,200 & 430,400,000 \\
LSTM                          & --                & [200, 10, 3]        & 660     &             \\ \bottomrule
\end{tabular}%
\end{table}

\subsubsection{Models training and hyperparameters}
For every model evaluation, we used 90\% and 10\% of the training dataset for training and validation, respectively. The validation dataset was used for early stopping after 20 epochs without improvement. All the models used a learning rate of 0.001, weight decay of 0.01, batch size of 200, and ADAM optimizer \cite{loshchilov2018decoupled}. To train DL models we used PyTorch \cite{NEURIPS2019}, skorch \cite{skorch}, and braindecode \cite{schirrmesiter_eeg}.

\subsection{Offline experiments}
First, we computed results in a classical scenario, i.e., train/valid/test split. We used the calibration dataset (first six sessions) as the training dataset. The rest of the data (online evaluation dataset) was used as the test set.

Additionally, we gradually increased the training dataset size from one session up to 22 with a step of 2. As different models may have different dataset requirements, we wanted to verify whether collecting more data can be more profitable for one of the evaluated optimization/architecture combinations.

To investigate the possible influence of end-to-end learning on models' robustness against data mislabelling, we perturbed the dataset to make training more challenging. In the BCI, part of observations is often mistakenly labeled due to lack of subject attention, tiredness, experimental setup, etc. Therefore, we randomly selected a fraction of observations in which targets were shuffled between samples so they no longer have a connection with the ECoG signal while preserving the same distribution. At the same time, we kept the test set unchanged.


\section{Results}

\begin{table}
\centering
\caption{Cosine similarity computed in the train-valid-test split scenario. Values are sorted by average performance and represent the mean and standard deviation of 5 runs.}
\label{tab:cs_tt}
\setlength{\tabcolsep}{6pt}
\begin{tabular}{lcc}
\toprule
{} &        Left hand &       Right hand \\
\midrule
E2E CNN+LSTM+MT CFO              &  $\mathbf{0.304\pm0.005}$ &  $0.266\pm0.020$ \\
CNN+LSTM+MT                      &  $0.297\pm0.008$ &  $0.270\pm0.011$ \\
E2E CNN+LSTM+MT                  &  $0.289\pm0.007$ &  $\mathbf{0.273\pm0.015}$ \\
E2E MLP CFO                      &  $0.254\pm0.012$ &  $0.230\pm0.013$ \\
MLP                              &  $0.247\pm0.023$ &  $0.232\pm0.005$ \\
E2E MLP                          &  $0.243\pm0.014$ &  $0.234\pm0.020$ \\
ShallowFBCSPNet \cite{schirrmesiter_eeg}                 &  $0.235\pm0.010$ &  $0.236\pm0.011$ \\
E2E CNN+LSTM+MT random init      &  $0.216\pm0.008$ &  $ 0.230\pm0.020$ \\
E2E MLP random init              &  $0.181\pm0.029$ &  $ 0.223\pm0.008$ \\
Deep4Net \cite{schirrmesiter_eeg}                     &  $0.111\pm0.021$ &  $0.259\pm0.013$ \\
\bottomrule
\end{tabular}
\end{table}

We started the analysis by comparing different model training scenarios when trained on the first six sessions (online calibration dataset). The results for the train/test split can be found in table \ref{tab:cs_tt}. Differences between scenarios are rather small, with only small performance improvement coming from full end-to-end optimization. The best performance was achieved by models using CFO. However, the gap between the hand-crafted features approach and CFO is relatively small, considering standard deviations of the computed values. The worst performance was achieved for Deep4Net (especially low performance for the left hand dataset) and both MLP and CNN+LSTM+MT models with random weights initialization, suggesting the high importance of the prior signal processing knowledge used to define the wavelet shape of the filters at the beginning of the training.

\begin{figure}
\centering
\includegraphics[width=.9\textwidth]{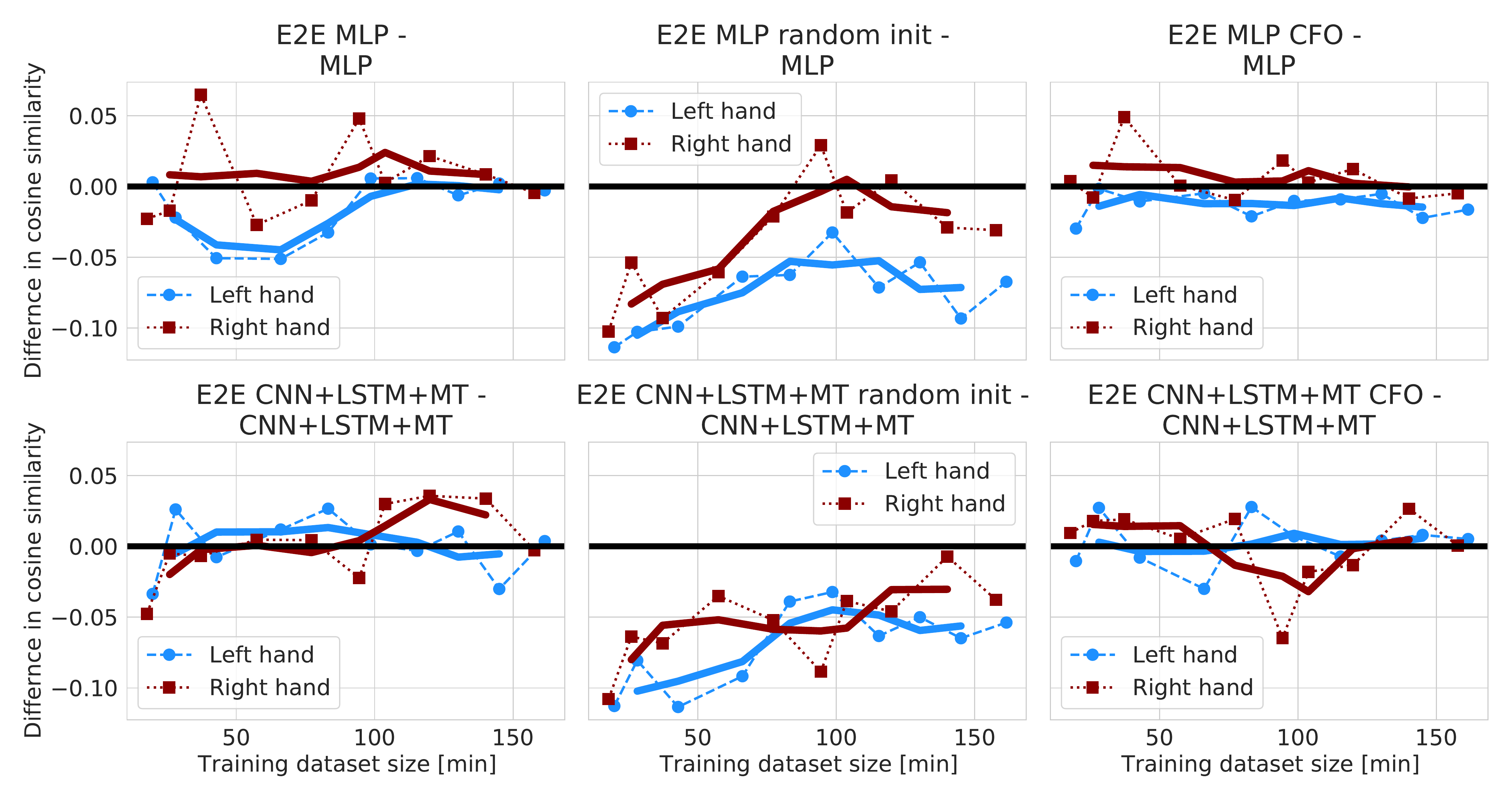}
\caption{Difference between end-to-end model and its counterpart using hand-crafted features. The bold line denotes the moving average with a window of size 3.} \label{fig:difference-dataset-size}
\end{figure}
We did not notice significant improvements coming from end-to-end optimization, so we wanted to verify the hypothesis of different dataset size requirements for different optimization methods. Therefore, the differences between end-to-end models and their hand-crafted features counterparts for several training dataset sizes are presented in figure \ref{fig:difference-dataset-size}. In some cases, end-to-end models increase the cosine similarity faster than the models using fixed features, so the gap between models can be reduced for approaches using random weights initialization. However, only for models initialized to wavelets and optimized directly, an improvement over hand-crafted features can be observed for some points (up to 0.05 of cosine similarity for the right hand dataset).

\begin{figure}
\centering
\includegraphics[width=.8\textwidth]{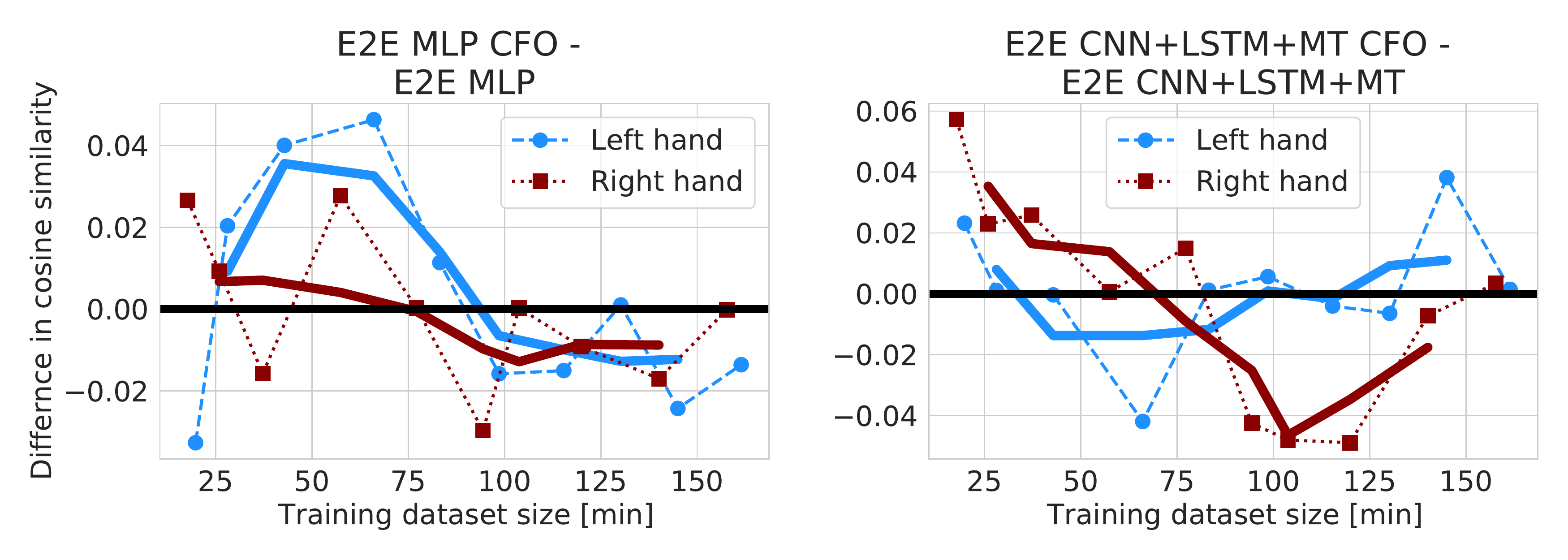}
\caption{Difference between the CFO model and its counterpart using constraint-free end-to-end optimization. The bold line denotes the moving average with a window of size 3.} \label{fig:difference-dataset-size-cfo-e2e}
\end{figure}

When comparing CFO and standard E2E optimization in figure \ref{fig:difference-dataset-size-cfo-e2e}, higher effectiveness of CFO for small training datasets can be observed. CFO may limit overfitting as the functions represented by the convolutional filters are constrained to the wavelet family. It may be interpreted as an additional optimization constraint imposed on model parameters. Note that diminished gap between CFO and standard end-to-end in figure \ref{fig:difference-dataset-size-cfo-e2e} show only relative decrease of CFO performance.

\subsection{Filters visualization}
\begin{figure}
    \centering
    \includegraphics[width=\textwidth]{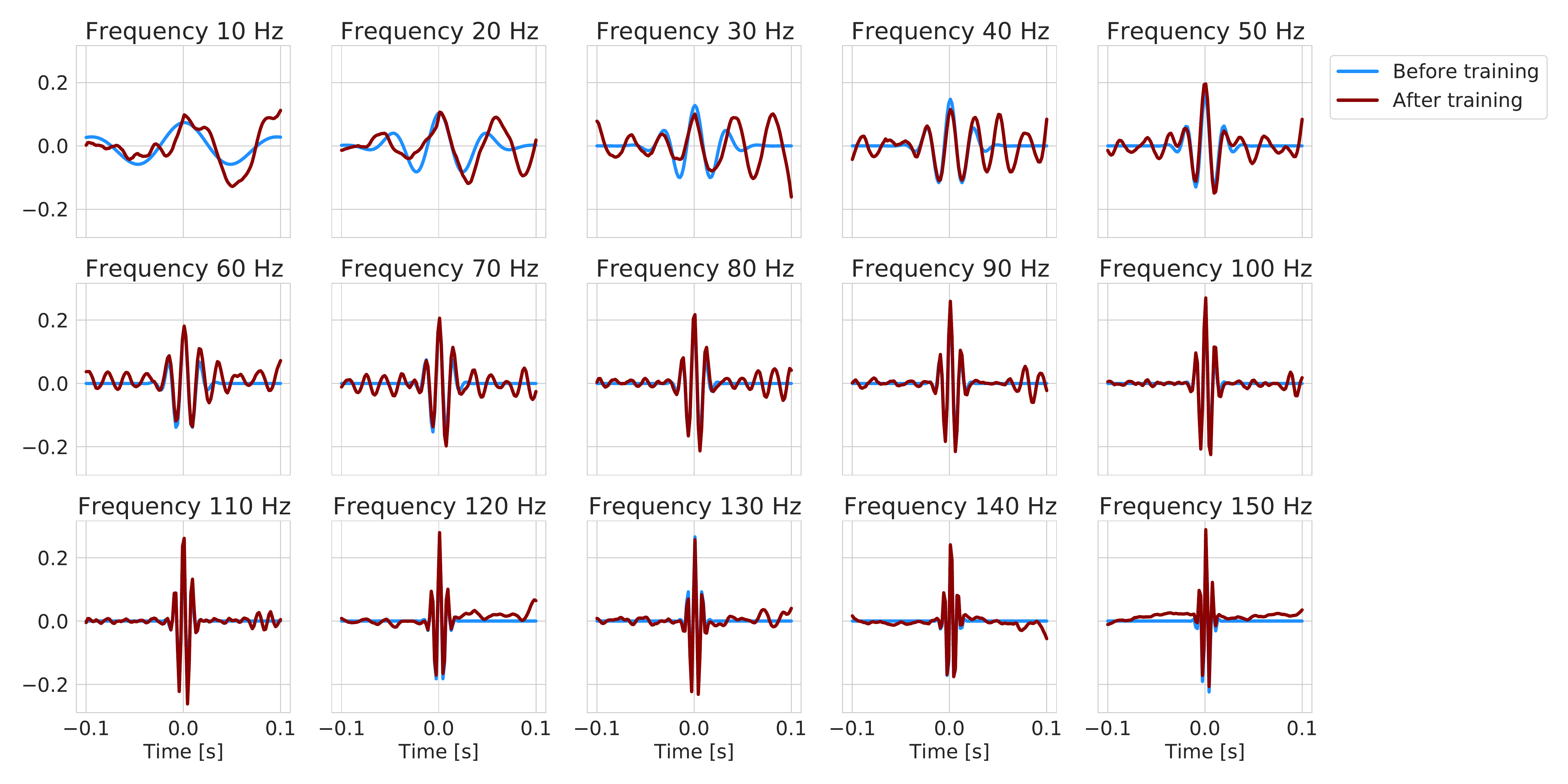}
    \caption{Visualized filters before (blue) and after (orange) training for the models with parameters optimized freely. Note that only real part of the wavelet was visualized for clarity. Plot titles denote central wavelet frequency at initialization.}
    \label{fig:wavelet-modification}
\end{figure}

We visualized the filters before and after training to analyze the characteristics of learned feature extraction. In figure \ref{fig:wavelet-modification}, we presented the filters modified without additional constraints. The biggest change can be observed in the central frequencies between 30 Hz and 80 Hz. In most cases, the initial central frequency was maintained, while the wavelets got extended with a signal similar to the sine wave in the central wavelet frequency. This could indicate the importance of information about frequencies from which the signal is composed. At the same time, extending wavelets reduces the temporal resolution of the signals. The changes in the high-frequency wavelets ($>100$Hz) are less significant, and the pattern of extending wavelets is no longer visible. Instead, components of significantly lower frequencies and smaller amplitude were added.

\begin{figure}
    \centering
    \includegraphics[width=\textwidth]{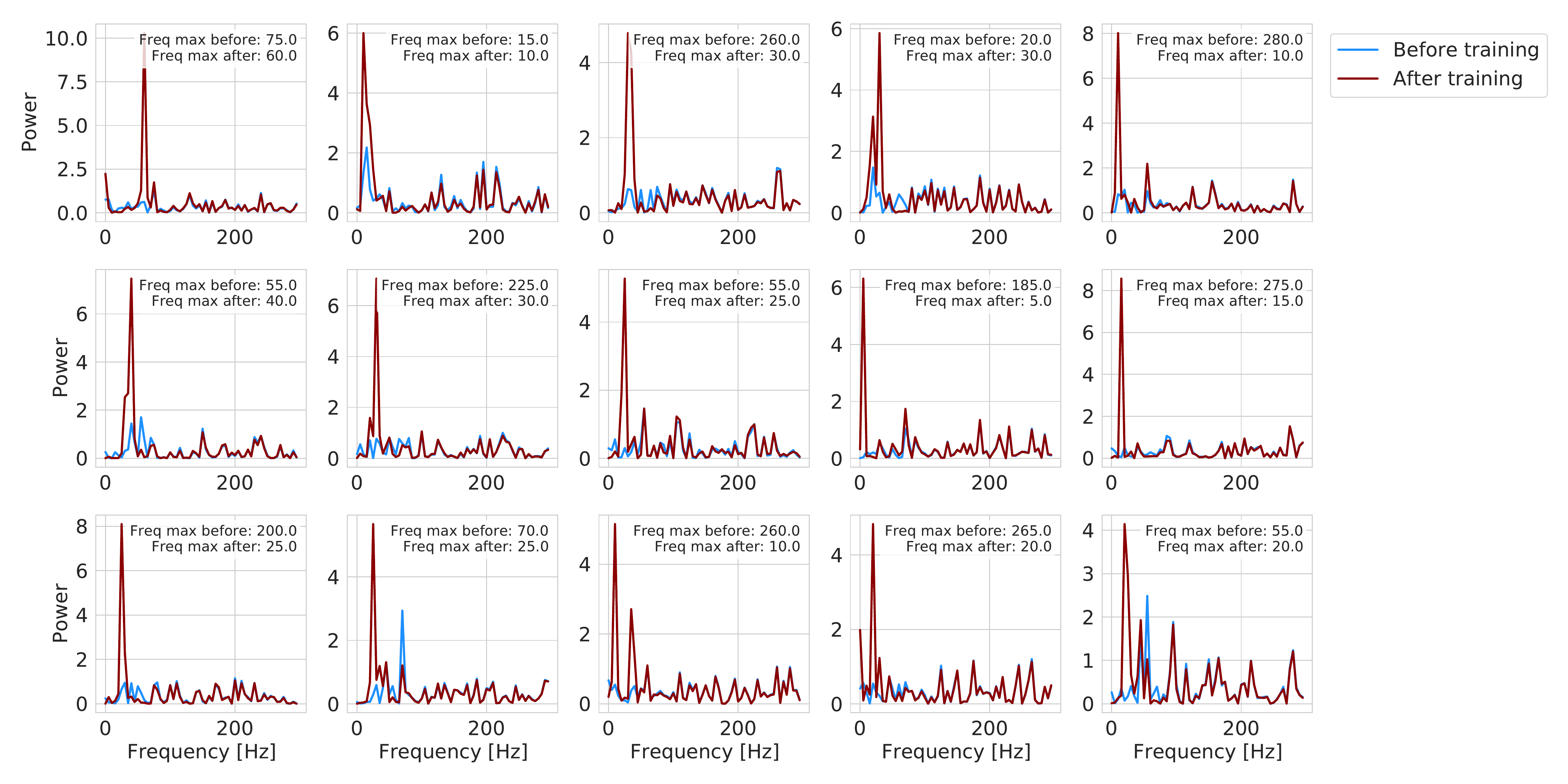}
    \caption{Power spectra of filters before (blue) and after (orange) training for convolutional layer initialized to random. The plots denoted frequencies for which maximum power spectra were observed before and after training.}
    \label{fig:random-modification}
\end{figure}

In figure \ref{fig:random-modification}, we visualized filters before and after optimization when the convolutional layer was initialized to random. As random filters were much harder to analyze visually, we presented them in the form of power spectra, so the changes in the filtered frequencies could be better visible. All filters have a maximum power peak lower than 65 Hz with 40\% of maxima contained in the frequency range 25-30Hz. Compared to hand-crafted features, end-to-end filters initialized to random covered only approximately half of the frequency band analyzed by the traditional feature extraction pipeline. However, in the higher frequencies, there are smaller peaks which can also contribute to the extracted representation and may cover the missing frequency band.

\begin{figure}
    \centering
    \subfloat[\centering standard central frequencies in range 10-150]{{\includegraphics[width=.47\textwidth]{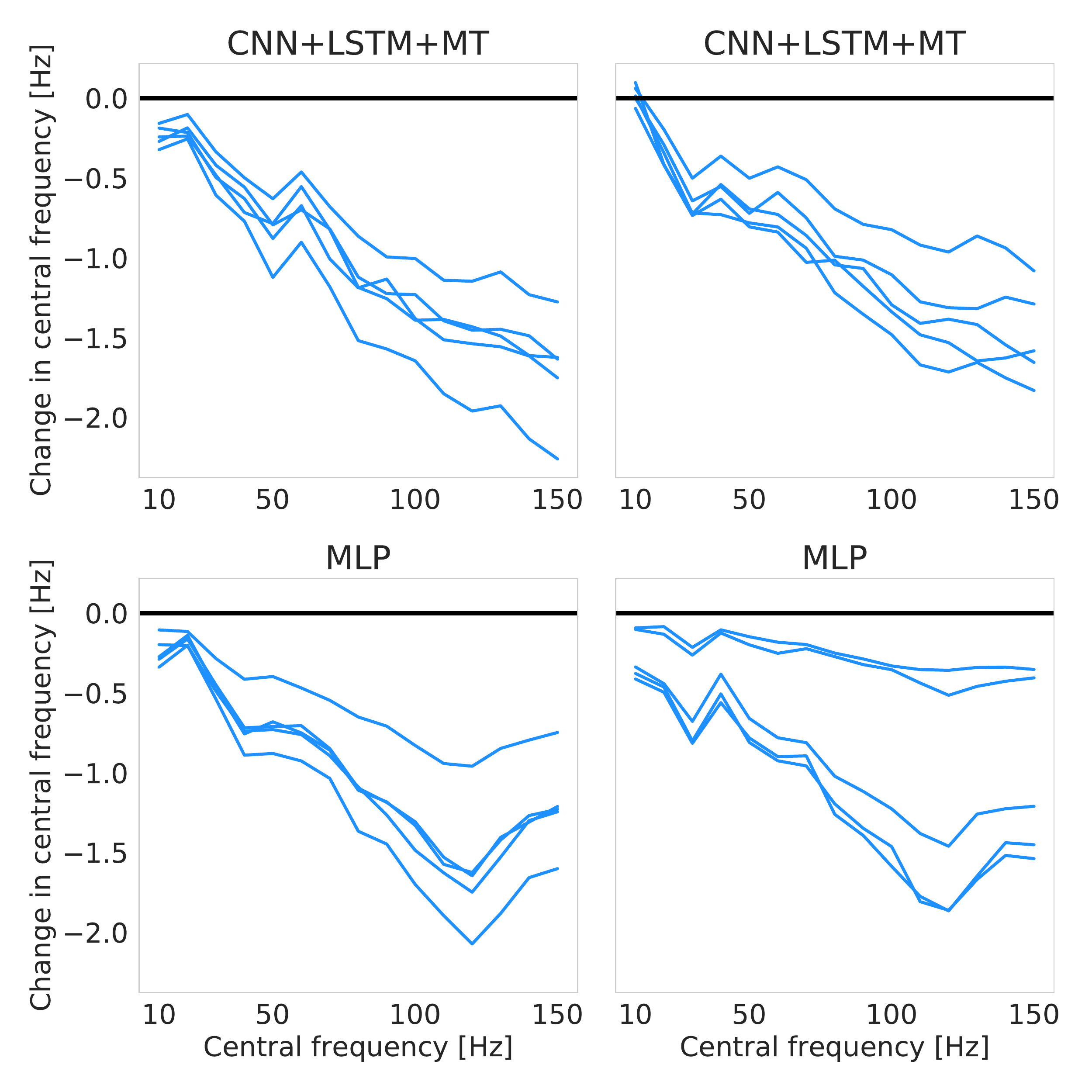}}}%
    \qquad
    \subfloat[\centering central frequencies squeezed to range 0-1]{{\includegraphics[width=.47\textwidth]{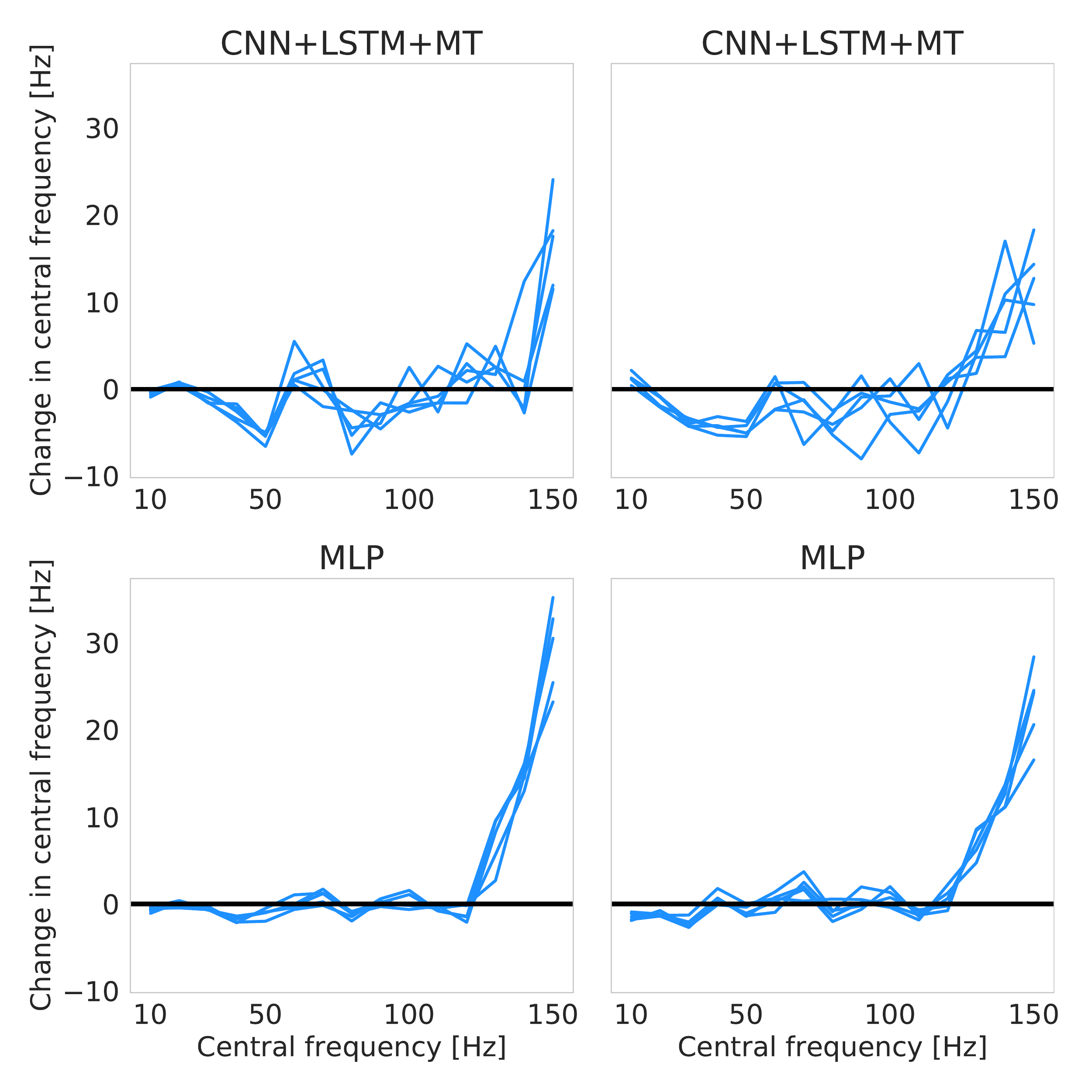}}}%
    \caption{Difference between central wavelet frequencies before and after CFO. Models for left hand translation are presented in the left column, for the right hand in the right column. Note that the scale is different for the (a) and (b) figures.}%
    \label{fig:cfo-modification}%
\end{figure}

In figure \ref{fig:cfo-modification}.a, we presented the difference between initialized central wavelet frequency and the one obtained after the training. We observed a decrease in almost all frequencies when training the models. The decrease was higher for higher frequencies. This may suggest that more information can be extracted from lower frequencies. However, in our preliminary results, we noticed that adapting the learning rate for the convolutional layer may significantly change the frequency behavior (see figure \ref{fig:cfo-modification}.b), which should be taken into account when analyzing the results. This may lead to different changes in the central frequencies than in the base model. 
The gradient was increased 150 times by squeezing central frequencies from 10-150Hz to 0-1. In the case of initialization to wavelet, a network may start the training near a local minimum found by the manual design of feature extraction that can be hard to move out. Setting a higher learning rate may enable reaching different regions on the loss function surface. The performance achieved with a higher learning rate was similar to the standard CFO results with a cosine similarity of $0.283\pm0.014$ (left hand) and $0.270\pm0.011$ (right hand) for CNN+LSTM+MT and $0.262\pm0.01$ (left hand) and $0.227\pm0.007$ (right hand) for MLP.

\subsection{Target perturbation}
\begin{figure}
    \centering
    \includegraphics[width=.8\textwidth]{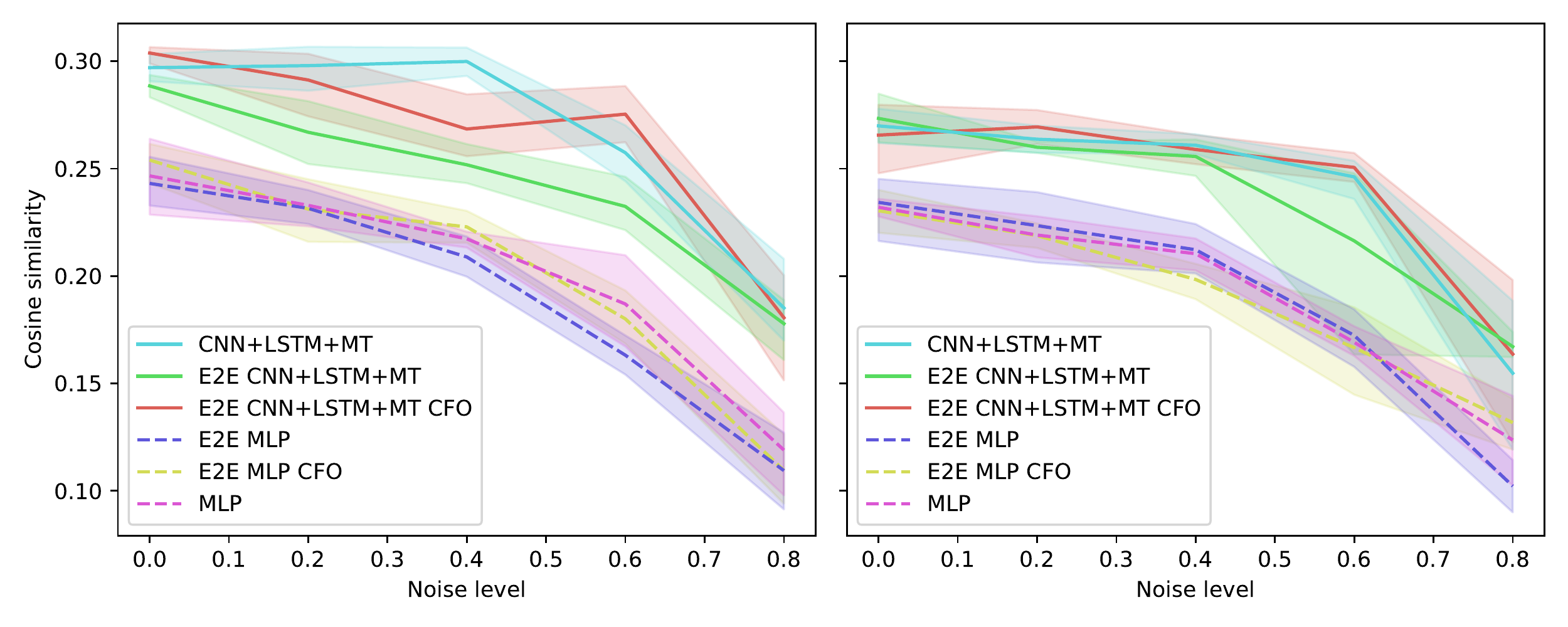}
    \caption{Influence of noise in the targets on models' performance. Noise level indicates the fraction of observations with perturbed labels.}
    \label{fig:target-noise}
\end{figure}

In the case of perturbed ground-truth (figure \ref{fig:target-noise}), CNN+LSTM+MT models were more robust to noise in the targets with increased stability (especially for the left hand) of hand-crafted features and CFO models compared to models optimized freely. On the other hand, in the case of MLP models, almost no differences between different optimization methods in the influence of noise on the performance were noticed.

\section{Discussion}
We proposed several approaches for the end-to-end optimization of deep learning ECoG decoders. However, in this study, we did not observe improvement from end-to-end optimization, especially when no prior knowledge was used for filter initialization. This confirms the usefulness of hand-crafted features and years of neuroscientific signal processing while leaving doors open to more sophisticated end-to-end models. Firstly, deeper models with more advanced DL mechanisms \cite{eeg_inception,eeg_transformer} should be evaluated as they may allow for the extraction of more complex representations and thus outperform hand-crafted features. Secondly, machine learning methods for robust learning may be evaluated, e.g., learning from noisy input data, noisy labels, and out-of-distribution samples \cite{li_noise}. Those can particularly tackle problems coming from specific recording/experimental circumstances.

The reasoning behind our study is focused on the specificity of ECoG brain signals and the adequacy of selected DL methods to the problem. The specificity originates from experimental constraints caused by the presence of a human in the loop but also signals characteristics, hardware capabilities, etc. It results in a distorted dataset with a low signal-to-noise ratio, short signal stationarity interval, and uncertain labels. This is quite different from computer vision problems, usually with well-defined labels and images understandable with a naked eye. Improving information extraction from noisy data may be especially important in the light of increased robustness to noise in targets shown by the CNN+LSTM+MT model compared to MLP. Using all 10 targets recorded during a 1-second window decreases the influence of single perturbed points on the performance because the information can be efficiently extracted even for 40\% or 60\% of perturbed targets. In this case, the CNN+LSTM+MT model using hand-crafted features maintains high performance for a higher noise level than the end-to-end model. However, an important point in the discussion is that our dataset, even after data cleaning, still contains a significant amount of observations with incorrect labels. Thus, in figure \ref{fig:target-noise}, a noise level equal to zero corresponds to an unknown noise level in labels originating from the experimental setup. Thus, generative models should be used to create datasets with a known level of noise and analyze the influence of perturbations on the performance in the case of less distorted datasets.

All the results were computed offline on datasets recorded with only one patient. These kinds of datasets are hardly accessible due to experimental and legal constraints. It makes the generalization of the results to other patients and datasets hard to estimate. Thus, more simulations should be performed to confirm our conclusions, ideally with more patients and tasks. This should also include hyperparameters search, like learning rate, batch size, weight decay, as those could vary between different approaches. However, performing hundreds of evaluations is time-consuming, and the problem is magnified in the case of end-to-end models due to increased computational load. Our study focused on feature extraction based on wavelet transform, which was previously used in this problem. As we optimized the parameters of the wavelet transform without changing other parts of the model, we isolated the influence of end-to-end optimization on models' performance. While this simplified the problem, our study did not evaluate other feature extraction pipelines that could behave differently. Thus, an extended analysis of several feature extraction pipelines compared to their end-to-end counterparts would allow for broader generalization and therefore is of great interest.

While this article and \cite{xie_decoding_2018} analyzed ECoG signals, targets used for training models in \cite{xie_decoding_2018} were actual fingers trajectories recorded while subjects performed real movements. In our case, targets are much noisier due to the lack of labeling based on the hand movements of a tetraplegic patient. This may favor hand-crafted features, as could be seen for CNN+LSTM+MT in figure \ref{fig:target-noise}. Finally, our conclusions are in line with \cite{xie_decoding_2018} who observed relatively small improvement from optimizing hand-crafted features and worse performance/longer training time when initializing the model to random. In our case, end-to-end models achieved the same performance as models using CWT features only with smart weights initialization, which emphasizes the importance of prior signal processing knowledge in designing DL for ECoG analysis.


\section*{Acknowledgements}
Clinatec is a Laboratory of CEA-Leti at Grenoble and has statutory links with the University Hospital of Grenoble (CHUGA) and University Grenoble Alpes (UGA). This study was funded by CEA (recurrent funding) and the French Ministry of Health (Grant PHRC-15-15-0124), Institut Carnot, Fonds de Dotation Clinatec. Matthieu Martin was supported by the cross-disciplinary program on Numerical Simulation of CEA. Maciej Śliwowski was supported by the CEA NUMERICS program, which has received funding from European Union's Horizon 2020 research and innovation program under the Marie Sklodowska-Curie grant agreement No 800945 — NUMERICS — H2020-MSCA-COFUND-2017.

\bibliographystyle{splncs04}
\bibliography{mybibfile}

\begin{thebibliography}{10}
\providecommand{\url}[1]{\texttt{#1}}
\providecommand{\urlprefix}{URL }
\providecommand{\doi}[1]{https://doi.org/#1}

\bibitem{benabid_exoskeleton_2019}
Benabid, A.L., Costecalde, T., Eliseyev, A., Charvet, G., Verney, A., Karakas,
  S., Foerster, M., Lambert, A., Morinière, B., Abroug, N., Schaeffer, M.C.,
  Moly, A., Sauter-Starace, F., Ratel, D., Moro, C., Torres-Martinez, N.,
  Langar, L., Oddoux, M., Polosan, M., Pezzani, S., Auboiroux, V., Aksenova,
  T., Mestais, C., Chabardes, S.: An exoskeleton controlled by an epidural
  wireless brain–machine interface in a tetraplegic patient: a
  proof-of-concept demonstration. The Lancet Neurology  \textbf{18}(12),
  1112--1122 (Dec 2019). \doi{10.1016/S1474-4422(19)30321-7}, number: 12

\bibitem{bert}
Devlin, J., Chang, M., Lee, K., Toutanova, K.: {BERT:} pre-training of deep
  bidirectional transformers for language understanding. In: Burstein, J.,
  Doran, C., Solorio, T. (eds.) Proceedings of the 2019 Conference of the North
  American Chapter of the Association for Computational Linguistics: Human
  Language Technologies, {NAACL-HLT} 2019, Minneapolis, MN, USA, June 2-7,
  2019, Volume 1 (Long and Short Papers). pp. 4171--4186. Association for
  Computational Linguistics (2019). \doi{10.18653/v1/n19-1423}

\bibitem{eliseyev_recursive_2017}
Eliseyev, A., Auboiroux, V., Costecalde, T., Langar, L., Charvet, G., Mestais,
  C., Aksenova, T., Benabid, A.L.: Recursive {Exponentially} {Weighted} {N}-way
  {Partial} {Least} {Squares} {Regression} with {Recursive}-{Validation} of
  {Hyper}-{Parameters} in {Brain}-{Computer} {Interface} {Applications}.
  Scientific Reports  \textbf{7}(1),  16281 (Dec 2017).
  \doi{10.1038/s41598-017-16579-9}, number: 1

\bibitem{NIPS2012_krizhevsky}
Krizhevsky, A., Sutskever, I., Hinton, G.E.: Imagenet classification with deep
  convolutional neural networks. In: Pereira, F., Burges, C., Bottou, L.,
  Weinberger, K. (eds.) Advances in Neural Information Processing Systems.
  vol.~25. Curran Associates, Inc. (2012),
  \url{https://proceedings.neurips.cc/paper/2012/file/c399862d3b9d6b76c8436e924a68c45b-Paper.pdf}

\bibitem{Lawhern_2018}
Lawhern, V.J., Solon, A.J., Waytowich, N.R., Gordon, S.M., Hung, C.P., Lance,
  B.J.: {EEGNet}: a compact convolutional neural network for {EEG}-based
  brain{\textendash}computer interfaces. Journal of Neural Engineering
  \textbf{15}(5),  056013 (jul 2018). \doi{10.1088/1741-2552/aace8c}

\bibitem{eeg_transformer}
Lee, Y.E., Lee, S.H.: Eeg-transformer: Self-attention from transformer
  architecture for decoding eeg of imagined speech (2021).
  \doi{10.48550/ARXIV.2112.09239}

\bibitem{li_noise}
Li, J., Xiong, C., Hoi, S.C.: Learning from noisy data with robust
  representation learning. In: 2021 IEEE/CVF International Conference on
  Computer Vision (ICCV). pp. 9465--9474 (2021).
  \doi{10.1109/ICCV48922.2021.00935}

\bibitem{liang_decoding_2012}
Liang, N., Bougrain, L.: Decoding {Finger} {Flexion} from {Band}-{Specific}
  {ECoG} {Signals} in {Humans}. Frontiers in Neuroscience  \textbf{6} (2012).
  \doi{10.3389/fnins.2012.00091}

\bibitem{loshchilov2018decoupled}
Loshchilov, I., Hutter, F.: Decoupled weight decay regularization. In:
  International Conference on Learning Representations (2019),
  \url{https://openreview.net/forum?id=Bkg6RiCqY7}

\bibitem{mestais_wimagine_2015}
{Mestais}, C.S., {Charvet}, G., {Sauter-Starace}, F., {Foerster}, M., {Ratel},
  D., {Benabid}, A.L.: Wimagine: Wireless 64-channel ecog recording implant for
  long term clinical applications. IEEE Transactions on Neural Systems and
  Rehabilitation Engineering  \textbf{23}(1),  10--21 (2015).
  \doi{10.1109/TNSRE.2014.2333541}

\bibitem{NEURIPS2019}
Paszke, A., Gross, S., Massa, F., Lerer, A., Bradbury, J., Chanan, G., Killeen,
  T., Lin, Z., Gimelshein, N., Antiga, L., Desmaison, A., Kopf, A., Yang, E.,
  DeVito, Z., Raison, M., Tejani, A., Chilamkurthy, S., Steiner, B., Fang, L.,
  Bai, J., Chintala, S.: Pytorch: An imperative style, high-performance deep
  learning library. In: Advances in Neural Information Processing Systems 32,
  pp. 8024--8035. Curran Associates, Inc. (2019),
  \url{http://papers.neurips.cc/paper/9015-pytorch-an-imperative-style-high-performance-deep-learning-library.pdf}

\bibitem{Roy_2019}
Roy, Y., Banville, H., Albuquerque, I., Gramfort, A., Falk, T.H., Faubert, J.:
  Deep learning-based electroencephalography analysis: a systematic review.
  Journal of Neural Engineering  \textbf{16}(5),  051001 (aug 2019).
  \doi{10.1088/1741-2552/ab260c}

\bibitem{eeg_inception}
Santamaría-Vázquez, E., Martínez-Cagigal, V., Vaquerizo-Villar, F., Hornero,
  R.: Eeg-inception: A novel deep convolutional neural network for assistive
  erp-based brain-computer interfaces. IEEE Transactions on Neural Systems and
  Rehabilitation Engineering  \textbf{28}(12),  2773--2782 (2020).
  \doi{10.1109/TNSRE.2020.3048106}

\bibitem{schalk_decoding_2007}
Schalk, G., Kubánek, J., Miller, K.J., Anderson, N.R., Leuthardt, E.C.,
  Ojemann, J.G., Limbrick, D., Moran, D., Gerhardt, L.A., Wolpaw, J.R.:
  Decoding two-dimensional movement trajectories using electrocorticographic
  signals in humans. Journal of Neural Engineering  \textbf{4}(3),  264--275
  (Sep 2007). \doi{10.1088/1741-2560/4/3/012}, number: 3

\bibitem{schirrmesiter_eeg}
Schirrmeister, R.T., Springenberg, J.T., Fiederer, L.D.J., Glasstetter, M.,
  Eggensperger, K., Tangermann, M., Hutter, F., Burgard, W., Ball, T.: Deep
  learning with convolutional neural networks for eeg decoding and
  visualization. Human Brain Mapping  \textbf{38}(11),  5391--5420 (2017).
  \doi{https://doi.org/10.1002/hbm.23730}

\bibitem{sliwowski_2022}
{\'{S}}liwowski, M., Martin, M., Souloumiac, A., Blanchart, P., Aksenova, T.:
  Decoding {ECoG} signal into 3d hand translation using deep learning. Journal
  of Neural Engineering  \textbf{19}(2),  026023 (mar 2022).
  \doi{10.1088/1741-2552/ac5d69}

\bibitem{Tabar_2016}
Tabar, Y.R., Halici, U.: A novel deep learning approach for classification of
  {EEG} motor imagery signals. Journal of Neural Engineering  \textbf{14}(1),
  016003 (nov 2016). \doi{10.1088/1741-2560/14/1/016003}

\bibitem{skorch}
Tietz, M., Fan, T.J., Nouri, D., Bossan, B., {skorch Developers}: skorch: A
  scikit-learn compatible neural network library that wraps PyTorch (Jul 2017),
  \url{https://skorch.readthedocs.io/en/stable/}

\bibitem{volkova_ecog_review}
Volkova, K., Lebedev, M.A., Kaplan, A., Ossadtchi, A.: Decoding movement from
  electrocorticographic activity: A review. Frontiers in Neuroinformatics
  \textbf{13} (2019). \doi{10.3389/fninf.2019.00074}

\bibitem{xie_decoding_2018}
Xie, Z., Schwartz, O., Prasad, A.: Decoding of finger trajectory from {ECoG}
  using deep learning. Journal of Neural Engineering  \textbf{15}(3),  036009
  (Jun 2018). \doi{10.1088/1741-2552/aa9dbe}, number: 3

\end{thebibliography}
\end{document}